\documentclass[prl,twocolumn,aps,showpacs,amsmath,amssymb]{revtex4}
\usepackage{graphicx}
\usepackage{dcolumn}
\usepackage{bm}

\begin{document}

\title{Majorana bound state in rotating superfluid
$^3$He-A between parallel plates}

\author{Y. Tsutsumi}
%\email{mizushima@mp.okayama-u.ac.jp}
\affiliation{Department of Physics, Okayama University,
Okayama 700-8530, Japan}
\author{T. Kawakami}
%\email{mizushima@mp.okayama-u.ac.jp}
\affiliation{Department of Physics, Okayama University,
Okayama 700-8530, Japan}
\author{T. Mizushima}
\affiliation{Department of Physics, Okayama University,
Okayama 700-8530, Japan}
\author{M. Ichioka}
\affiliation{Department of Physics, Okayama University,
Okayama 700-8530, Japan}
\author{K. Machida}
\affiliation{Department of Physics, Okayama University,
Okayama 700-8530, Japan}
\date{\today}

\begin{abstract}
A concrete and experimentally feasible example for testing 
the putative Majorana zero energy state bound in a vortex
is theoretically proposed for a parallel plate geometry of
superfluid $^3$He-A phase. We examine the experimental setup 
in connection with ongoing rotating cryostat experiments.
The theoretical analysis is based on the well-established Ginzburg--Landau functional, supplemented by microscopic
calculations of the Bogoliubov--de Gennes equation, both of which allow 
the precise location of the parameter regions of the 
Majorana state to be found in realistic situations.

\end{abstract}


%67.30.he	Textures and vortices
%67.30.ht	Restricted geometries
%71.10.Pm	Fermions in reduced dimensions
\maketitle

Much attention has been focused on Majorana zero-energy
states in various research fields\cite{dassarma}, for example, leptogenesis in cosmology
in connection with the matter--antimatter imbalance problem\cite{majorana},
the fractional quantum Hall state, chiral superconductors, and $p$-wave
superfluids in neutral atomic gases.
Of particular interest is the possible application to
quantum computing, based on the fact that a pair of Majorana
states is intrinsically entangled and topologically 
protected from external disturbance.
For chiral superconductors and $p$-wave superfluids,
the key is finding the Majorana zero energy state bound in a vortex.
Its existence is guaranteed by the index theorem\cite{dassarma},
implying in this context that in spinless chiral $p$-wave pairing $p_x\pm ip_y$
of two dimension (2D), an odd winding number vortex always possesses
a Majorana state at exactly zero energy relative to the Fermi level.
This state is localized at the core\cite{volovik}.

To date there is has been no experimental evidence to prove its existence
in either chiral superconductors\cite{machida} 
or $p$-wave superfluids in neutral atomic gases, 
as neither of these phenomena have yet been realized.
Therefore, concrete
examples of $p$-wave superfluidity are needed.
In this respect, the $^3$He-A phase is a prime candidate for testing the Majorana state.

The ABM (A) phase in superfluid $^3$He is characterized by the order parameter (OP)
$A_{\mu i}=\Delta_0 d_{\mu}({\vec n}+i{\vec m})_i$ ($i=x, y, z$).
The spin and orbital structure of $p$-wave pairing, 
which is generally defined as  $\hat{\Delta}=i\sum_{\mu i} \left( A_{\mu i} 
\sigma_{\mu} \hat{p}_i \right) \sigma_y$,
are characterized by 
the $\vec d$ vector and $\vec l$ vector, respectively\cite{leggett,vollhardt}. 
($\vec l=\vec n\times \vec m$,
$\vec m$ and $\vec n$ are unit vectors, forming a triad.)
It is known that the $\vec l$ vector always points perpendicular to a 
wall surface and $\vec d$ vector is perpendicular to the applied field direction
if its magnitude is sufficiently strong 
($H>H_d=2.0$ mT, where $H_d$ is the dipole field\cite{leggett}).
Therefore, we can control the pairing symmetry both by the confining geometry 
and the field. This kind of controllability in superfluid $^3$He
is a great advantage over other research fields, in addition
to the decisive fact that the precise OP forms of the $p$-wave pairing are
established for the ABM (A) and BW (B) phases\cite{leggett,vollhardt}.

We consider a parallel plate geometry with a 
gap comparable to the dipole coherent
length $\xi_d\sim 10$ $\mu$m and a field ($H>H_d$) applied along the plates.
In this configuration the two vectors $\vec l\parallel \vec d$ 
are locked together perpendicular 
to the plates to maximize the dipole and magnetic energies,
and hence the OP is $A_{\mu i}=\Delta_0d_{\mu}({\vec n}+i{\vec m})_i$
($i=x, y$) with $\mu$ being fixed.
This is a ``spinless'' chiral superfluid, 
a situation ideal for the Majorana zero energy state\cite{dassarma}.

This parallel plate geometry has been already realized \cite{hakonen,yamashita},
namely a $^3$He-A phase sample\cite{yamashita} confined 
in a thin cylindrical region with thickness 12.5 $\mu$m and
radius $R=1.5$ mm. This thin plate with a 
cylindrical region is stacked  to 
form a layered structure consisting of 110 plates.
In an effort to observe a half-quantum vortex, the structure was rotated up to
a rotation speed of $\Omega$=6.28 radian/sec (rad/sec) in a rotating cryostat
at ISSP, University of Tokyo.
The texture transition under rotation was monitored by
an NMR spectrum to identify each vortex.

The experimental geometry of this previous experiment is ideal for realizing
the Majorana zero energy state, and hence we have adopted it for our study.
The necessary conditions for the realization of the Majorana state are:
(1) The bulk superfluid has a chiral $p$-wave symmetry.
(2) The vortex is singular with an odd winding number.
Condition (1) is satisfied because 
the $^3$He-A phase is orbitally triplet. 
As mentioned, the $\vec d$ and $\vec l$ vectors are locked 
together perpendicular to the plane.
The superfluid state is in the chiral $p$-wave symmetry 
characterized by $p_{\pm}=\mp(p_x\pm ip_y)/\sqrt2$
with frozen spin freedom.

However, we must carefully examine condition (2). 
There are several possible vortex states under this configuration,
either singular or non-singular vortices,
and compete with each other energetically.
The Majorana zero energy state is supported only 
by the singular vortex in general.
Thus it is important to
determine the possible stable parameter space spanned 
by the rotation speed $\Omega$, temperature $T$,
radius $R$ of the system, and pressure $P$
in terms of realistic experimental conditions\cite{yamashita}.
We also study the microscopic excitation spectrum
of the Majorana state on the three-dimensional (3D) Fermi surface by
solving the microscopic Bogoliubov--de Gennes (BdG) equation 
for a vortex. Previous theoretical studies have been
limited to strictly 2D systems\cite{tewari}. 
This study clarifies the detailed spectral
structures in the bulk and at the boundary where the edge current flows,
i.e., the dispersion relation of the Majorana zero mode
in 3D.

We start with the following well-established Ginzburg--Landau (GL)
functional\cite{leggett,vollhardt} to examine the stable regions for the Majorana state.
This GL form is known to be adequate for describing 
the $^3$He-A phase where the OP is described by $A_{z i}=\hat{d}_{z} A_i$:

\begin{eqnarray}
f=f_{bulk}+f_{grad}
\end{eqnarray}
\begin{eqnarray}
f_{bulk}=-\alpha A_i^* A_i +\beta_{13} A_i^* A_i^* A_j A_j +\beta_{245} A_i^* A_j^* A_i A_j
\end{eqnarray}
\begin{eqnarray}
f_{grad}=-K\left[ A_i^* \partial_i \partial_j A_j+A_j^* \partial_i \partial_i A_j
+A_j^* \partial_i \partial_j A_i \right]
\end{eqnarray}

\noindent
where $\partial_i= \nabla_i -i\frac{2m_3}{\hbar }(\vec{\Omega} \times \vec{r})_i$, $m_3$ is the mass of a $^3$He atom, $K$ is the effective mass, $\alpha=\alpha_0 (1-T/T_c)$, $\beta_{13}=\beta_1+\beta_3$ and $\beta_{245}=\beta_2+\beta_4+\beta_5$\cite{leggett,vollhardt}. 
We introduce $A_{\pm}=\mp(A_x\mp iA_y)/\sqrt2$
with $A=A_ip_i=A_+p_++A_0p_0+A_-p_-$, $(p_z=p_0)$.
Here we take into account terms up to the fourth order invariants and drop the spin degrees of freedom. 
As mentioned, the parallel plate geometry has a narrow
gap such that the $\vec l$-vector is fixed normal to the plate surfaces, implying that $A_z=0$.

The free energy is minimized numerically for various radii $R$ where we employ
the boundary condition $A_i(x,y)=0$ at the circumference $R$.
The superfluid current density is given by $j_i=\frac{4m_3K}{\hbar} Im \left[ A_i^* \partial_j A_j +A_j^* \partial_i A_j +A_j^* \partial_j A_i \right]$.
We use known GL parameters appropriate for our settings\cite{Sauls:PRB24:183, Kita:PRB66:224515, Greywall:PRB33:7520}.
The employed GL parameters\cite{GLparameter} are for $P$ = 3.2 MPa,
with a sufficiently strong $H$; $H>>H_d$ ($H$ = 26.7 mT in \cite{yamashita}). 
These are precisely known and well documented thanks to the long history 
of this research field\cite{leggett,vollhardt}.
The external rotation vector $\vec\Omega$
is parallel to the cylindrical axis $z$.
For $t=T/T_c >0.75$ ($T_c$ = 2.4 mK) the A-phase is stable.

We now consider the stable state at rest.
Our OP is the chiral axial states $A_{\pm}$,
which are degenerate at $\Omega=0$. The external 
rotation removes this degeneracy. We denote a state's 
winding number combination as $(w_+, w_-)$, where 
$A_{\pm}=|A_{\pm}|e^{iw_{\pm}\theta}$ ($\theta$ is the azimuthal angle).
Under our convention for rotation sense
(counter clockwise rotation) the (0,2) state is stabilized over
the (2,0), that is, the dominant component is  $A_{+}$
with the winding $w_+=0$,
as shown in Fig. 1. $A_{-}$ with the winding $w_-=2$
is induced at the boundary because the gradient terms
such as $A_+\nabla^2A_-$ inevitably induce
other components at places where the main OP is spatially
varied, at the boundary and vortex core in our case.
Under counter clockwise rotation the mass current flows less for (0,2) than for (2,0).
Thus at rest and in lower rotation regions, (0,2) is the ground state.
In Fig. 1 we display the OP profiles. The main component $A_+$ has no
phase winding, while the minor component $A_-$ has $4\pi$ phase winding, which
leads to spontaneous mass flow around the outer boundary even at rest.

We now examine a stable vortex under rotation.
There are a variety of possible vortex states.
The general form of the winding number combination
in the axis symmetry confined in a cylinder is given by
$(w_+=n, w_-=n\pm2)$ with $n$ an integer, meaning that through the
gradient term, two components with $w_+=n$ and  $w_-=n\pm2$ are coupled.
This includes (-1,1), (1,-1), (0,2), (2,0), (1,3), (-3,-1), (2,4), etc.
Note that in general, vortices with higher winding number are less plausible.
The (-1,1) state is comparable in energy with the (1,3) vortex.
The former is more stable than the latter for systems with smaller radii, 
$R=2.5, 5.0, 10$ $\mu$m in the lower rotation region.
However, since the (2,0) state in that region has the lowest energy,
(-1,1) is not realized.
The (1,-1) vortex is another candidate, but it is not stable in our calculation.
The higher winding vortex states, such as (2,4) become more stable than
the (1,3) vortex at higher rotations.
From these arguments, which we have confirmed
by detailed numerical calculations, we are left with only one stable state
(1,3) next to (0,2) upon increasing rotation.
Figure 2 displays the spatial structures for 
two components and phase windings of the (1,3) vortex 
where the dominant and minor components
are $A_+$  and $A_-$ with $w_+=1$ and $w_-=3$, respectively.

For (0,2) versus (1,3), 
it is obvious that under rotation the energy gain for (0,2)
due to ${\vec j}\cdot{\vec \Omega}$ is less than 
that for (1,3), which leads eventually to a phase transition from
(0,2) at rest to a single vortex with (1,3) with increasing
$\Omega$.
We have performed extensive calculations to find the critical
rotation speed $\Omega_c$
as a function of $R$ that allows us to safely 
locate the parameter
regions in the 
realistic case of $R$ = 1.5 mm. 
A difficulty of the numerical calculations 
is that we must treat two quite distinct
length scales, the coherence length $\xi\sim10$ nm 
and the system size $R$=1.5 mm, on an equal footing in 
the numerics.
The possibility of a singular core of 
$\xi$ scale was neglected in previous calculations.
Here we have solved the full GL functional to take the singular core
into account, using fine mesh points in the numerics. Buchholtz and Fetter\cite{fetter}
pointed out the possibility of a singular vortex in a narrow channel.

We show the results of our calculations in Fig. 3,
showing the size $R$ dependence of the critical rotation
speed $\Omega_c$. We have computed various sizes
up to $R=100$ $\mu$m at a fixed temperature $t=T/T_c=0.95$.
The inset shows an example determination of $\Omega_c$ for $R=0.8$ $\mu$m.
It can be seen in the main panel in Fig. 3 that 
$\Omega_c$ = 0.0 6rad/sec for $t$=  0.95 at $R$ = 1.5 mm, 
estimated by an extrapolation.
It should be feasible to realize this relationship experimentally.

For the temperature dependence of $\Omega_c$ 
we have checked that the critical speed  $\Omega_c$ 
slightly increases with lowing $T$.
We also found that $\Omega_c$ delicately depends on
$R$. By using an appropriate $R$ we can tune $\Omega_c$ to
bring it to an experimentally available rotation region. 
A maximum rotation speed of $\Omega=12$ rad/sec 
can be achieved at ISSP, University of Tokyo.
Thus, according to Fig. 3, $R$ must be above $R>100$ $\mu$m.

\begin{figure}
\includegraphics[width=8cm]{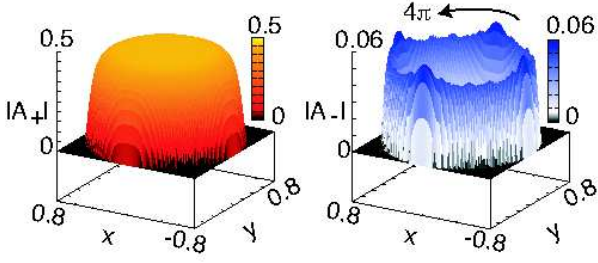}
\caption{(Color online) 
Spatial structures of the order parameters $|A_+|$ (left hand side)
and $|A_-|$ (right hand side) for the (0,2) state.
The sense of the phase winding and its number are shown.
$R=0.8$ $\mu$m and $\Omega=100$ rad/sec.
}
%\label{fig:egns}
\end{figure}

\begin{figure}
\includegraphics[width=8cm]{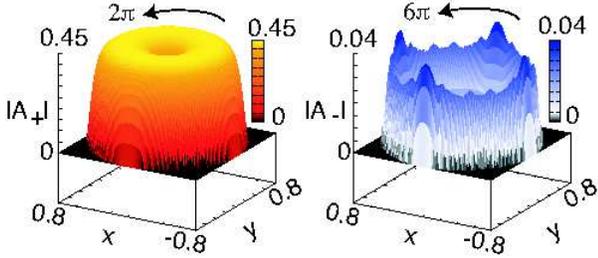}
\caption{(Color online) 
Spatial structures of the order parameters, $|A_+|$ (left hand side)
and $|A_-|$ (right hand side) for the (1,3) state.
$R=0.8$ $\mu$m and $\Omega=100$ rad/sec.
}
%\label{fig:egns}
\end{figure}

\begin{figure}[h!]
\includegraphics[width=7cm]{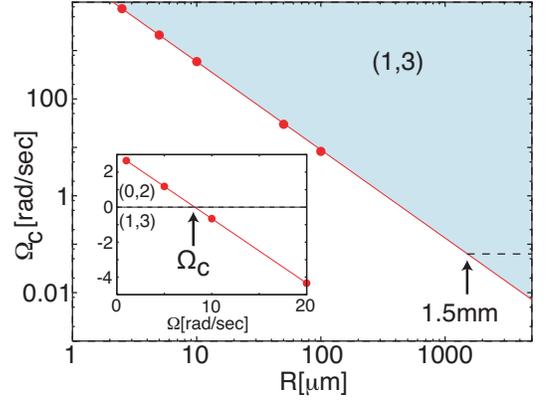}
\caption{(Color online)
Size $R$ dependence of the critical angular velocity
$\Omega_c$ from (0,2) at rest to (1,3) at a higher rotation
($T/T_c=0.95$).
An extrapolated value of $\Omega_c=0.06$ rad/sec
is found at $R=1.5$ mm.
The inset shows the  energy differences (arbitrary scale) between the two states as 
a function of $\Omega$ for $R=100$ $\mu$m.
}
%\label{fig:egns}
\end{figure}

Having found that the singular vortex characterized by $(1,3)$
is stabilized under rotation above $\Omega_c$, 
we analyze the spectral structure
of the core localized bound states in the $(1,3)$ vortex
with the 3D Fermi surface. 
The system is three dimensional.
We solve the following BdG equation under these conditions
to examine the Majorana zero energy state:
\begin{eqnarray}
\left[
\begin{array}{cc}
H({\bm r}) & \Pi({\bm r}) \\ - \Pi^{\ast}({\bm r}) & -H^{\ast}({\bm r})
\end{array}
\right]
\left[\begin{array}{c}
u_{\bm q}({\bm r}) \\ v_{\bm q}({\bm r}) 
\end{array}\right]
= E_{\bm q}
\left[\begin{array}{c}
u_{\bm q}({\bm r}) \\ v_{\bm q}({\bm r}) 
\end{array}\right], 
\label{eq:bdg}
\end{eqnarray}
where $\Pi ({\bm r}) \!=\! - \frac{i}{k_F} \sum _{\pm}\{ 
A_{\pm}({\bm r}) \square _{\pm} +\frac{1}{2}\square _{\pm}A_{\pm}({\bm r})
\}$ and $H({\bm r}) \!=\! -\frac{\hbar^2\nabla^2}{2m_3} -E_F - \Omega L_z$. We set $\square _{\pm} \!=\! \mp e^{\pm i \theta}(\frac{\partial}{\partial r} \pm \frac{i}{r}\frac{\partial}{\partial \theta})$ and $L_z \!\equiv\! -i\hbar\frac{\partial}{\partial \theta}$ in the cylindrical coordinates ${\bm r} \!=\! (r,\theta,z)$. Equation~(\ref{eq:bdg}) is self-consistently coupled with the gap equation 
$A_{\pm } ({\bm r}) \!=\! \frac{ig}{k _F} \sum _{|E_{\bm q}|\!<\! E_c}
[ v^{\ast}_{\bm q}({\bm r}) \square _{\mp} u_{\bm q}({\bm r}) 
- u_{\bm q}({\bm r})\square _{\mp} v^{\ast}_{\bm q}({\bm r}) ] f_{\bm q}$, 
where $f_{\bm q} \!=\! 1/(e^{E_{\bm q}/k_BT}+1)$ is the Fermi distribution function at a temperature $T$ and we henceforth set $T\!=\! 0$. The cutoff energy in the gap equation is set as $E_c \!=\! E_F$ and the coupling constant $g$ is chosen to be $k_F\xi \!=\! 2(\Delta/E_F)^{-1}\!=\! 14.3$ with $\Delta\!\equiv\! \max |A_{\pm}({\bm r})|$. 

We assume cylindrical symmetry of the OP having a single vortex at $r\!=\!0$, {\rm i.e.}, $A_{\pm}({\bm r}) \!=\! A_{\pm}(r) e^{i w_{\pm}\theta}$ with winding numbers $w_{+} \!\equiv\! n$ and $w_{-}\!\equiv\! n+2$. Also, we impose a periodic boundary condition with periodicity $Z \!=\! 250k^{-1}_F$ along the $z$ axis and a rigid boundary condition at $r \!=\! R\! =\! 250k^{-1}_F$ in Eq.~(\ref{eq:bdg}), which allows us to classify the eigenstates of Eq.~(\ref{eq:bdg}) with the azimuthal quantum number $q_{\theta} \!\in\! {\mathbb Z}$ and the linear quantum number $q_z \!=\! \frac{2\pi}{Z}n_z$ with $n_z\!\in\!\mathbb{Z}$. To this end, we find $u_{\bm q}({\bm r})\!=\!u_{\bm q}(r) e^{iq_{\theta}\theta}e^{iq_zz}$, $v_{\bm q}({\bm r})\!=\!v_{\bm q}(r) e^{i(q_{\theta}-n-1)\theta}e^{iq_zz}$. It is worth mentioning that for the $(1,3)$ vortex the eigenenergy under arbitrary $\Omega$ yields a linear shift from the value without rotation $E_{\bm q}(\Omega \!=\! 0)$ as 
\begin{eqnarray}
E_{\bm q} = E_{\bm q}(\Omega \!=\! 0) - \left(q_{\theta} - \frac{n+1}{2} \right)\Omega,
\label{eq:omega} 
\end{eqnarray}
under low rotation within $\Omega \!\ll\! E_F$.

A singular vortex produces a quantum well potential for quasiparticles around the core within $r\!\ll\! \xi$, which gives rise to a low energy eigenstate bounded at the core via Andreev scattering. Under no rotation, the eigenenergy of the core bound state in the 3D singular vortex with arbitrary winding $n$ can be analytically obtained from Eq.~(\ref{eq:bdg}) within $|q_{\theta}| \!\ll\! k_F\xi$ and $|q_{z}| \!\ll \! k_F$ \cite{mizushima}:
\begin{eqnarray}
E_{\bm q} = - \left( q_{\theta}-\frac{n+1}{2}\right) \omega _0
+ \left( m - \frac{n+1}{2} \right) \sin(\alpha) \omega _1, 
\label{eq:cdgm}
\end{eqnarray}
where $\sin^2(\alpha) \equiv\! 1 - \frac{q^2_{z}}{k^2_F}$ and $m\!\in\!\mathbb{Z}$. Here, assuming $A_{+}(r) \!=\! \Delta\tanh(r/\xi)$ and $A_{-}(r) \!=\! 0$, we find $\omega _0 \!\simeq\! - n \frac{\Delta^2_0}{E_F}$ and $|\omega _1| \!\sim\! \Delta$. The remarkable feature of Eq.~(\ref{eq:cdgm}) is that the eigenenergy can become exactly zero in a vortex of odd number $n$, in spite of the discreteness of the energy due to the narrow core. Also, the zero energy state is dispersionless on the linear momentum $q_z$. Note that the zero energy state appears at $q_{\theta} \!=\! 1$ in the $(1,3)$ vortex with $n\!=\! 1$ and at $q_{\theta} \!=\! 0$ in the $(-1,1)$ vortex with $n \!=\! -1$. It is found from Eq.~(\ref{eq:omega}) that the eigenenergy always lies at zero energy against the increase of the rotation frequency $\Omega$.

\begin{figure}[b!]
\includegraphics[width=8cm]{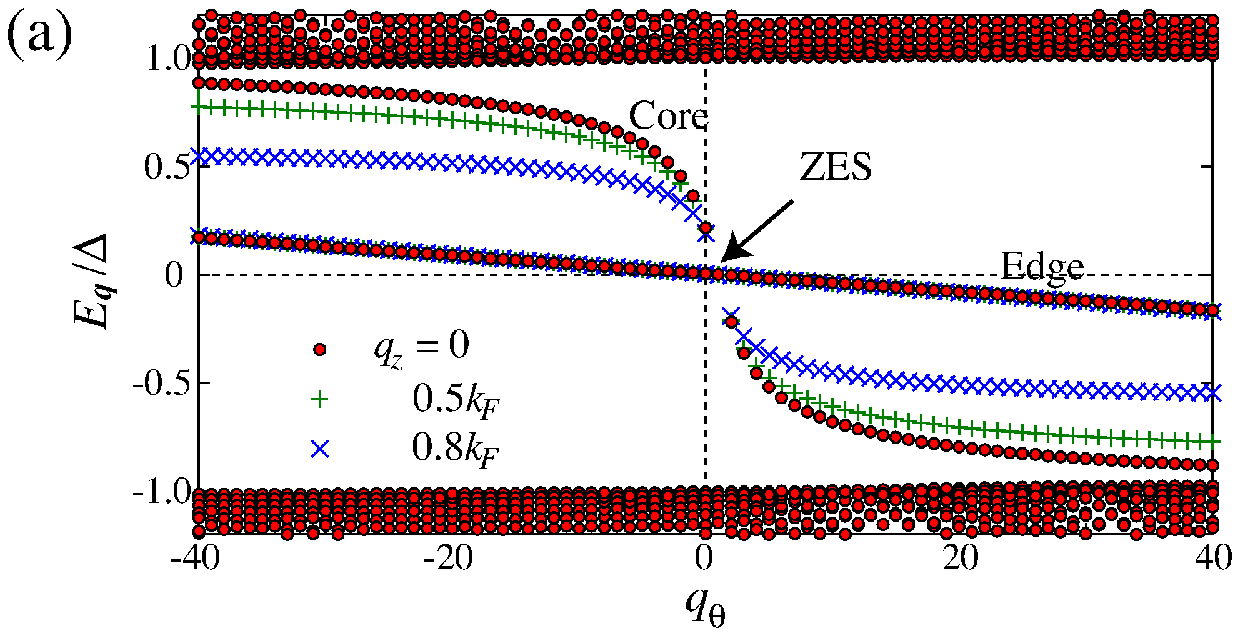}
\includegraphics[width=8cm]{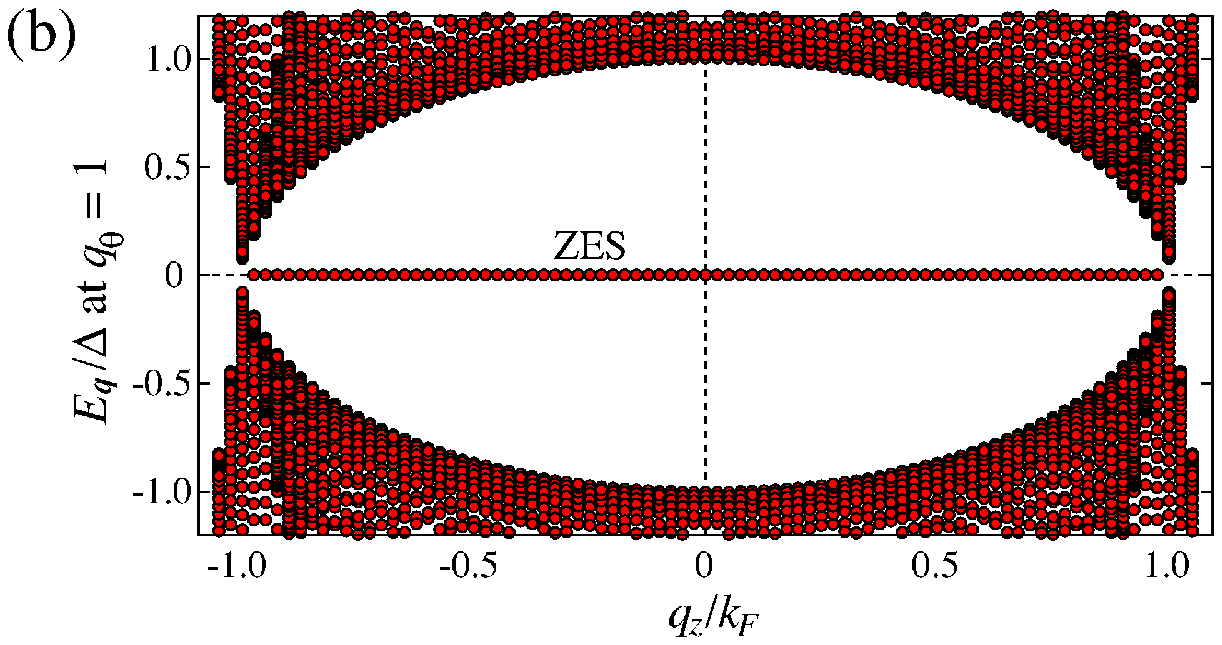}
\caption{(Color online) Energy spectrum $E_{\bm q}$ with $q_z\!=\!0$ (a) and $q_{\theta} \!=\! 1$ (b) in the $(1,3)$ vortex under no rotation $\Omega \!=\! 0$, where $\Delta \!\equiv\! \max|A_{\pm}(r)|$. The branches labeled ``Core'' and ``Edge'' in (a) are eigenstates having wave functions tightly bounded at the vortex core and the edge, respectively. The zero energy state (ZES) appears within $|q_{z}|\!<\! k_F$ at $q_{\theta} \!=\! 1$.
}
\label{fig:egns}
\end{figure}

To clarify the stability of the Majorana zero energy state, we present in Fig.~\ref{fig:egns} the low-lying quasiparticle states of the $(1,3)$ vortex as functions of 
$q_{\theta}$ and $q_z$, obtained from a full numerical calculation of Eq.~(\ref{eq:bdg}) along with the gap equation. Note that $|A_{+}(r)| \!\gg\! |A_{-}(r)|$. It can be seen that the zero energy state appears at $q_{\theta}\!=\! 1$, which is dispersionless situated exactly at $E=0$ within $|q_z| \!<\! k_F$. In Fig.~\ref{fig:egns}(b), the fact that the bulk excitation gap changes from the maximum value $|E_{\bm q}| \!=\! \Delta$ at $q_{z} \!=\! 0$ to $|E_{\bm q}| \!\sim\! 0$ at $q_z \!=\! \pm k_F$ indicates the presence of point nodes, due to the A phase,
situated at the north and south poles on the 3D Fermi surface. Since the zero energy state results from the resonance between two low-lying modes, such as the core bound and edge modes, it yields two-fold degeneracy and its wave function is localized at both the vortex core and edge, where $u_{\bm q}({\bm r}) \!\sim\! k_Fr$ inside the core region $r\!\ll\!\xi$. It should be emphasized again that the flat band structure of the zero energy states along $q_{z}$ is independent of the rotation frequency $\Omega$, because the angular momentum of the quasiparticles $q_{\theta} \!=\! 1$ is canceled by the vorticity $+1$ and the chirality $+1$ of the Cooper pair, as seen in Eq.~(\ref{eq:omega}). 

The Majorana core bound state
and the edge mode inevitably hybridize, causing
level splitting when the system is small enough.
For the detection of the Majorana state, an experiment could be devised to
search for the energy splitting, which
is absent in ordinary vortex bound
state. The energy splitting could be detected by sound attenuation
or electromagnetic absorption experiments.

It is possible to study similar Majorana physics for a
different configuration with the field orientation perpendicular to the plates, as is
realized in Ref.~\cite{yamashita}.
Since the $\vec d$-vector is now free to rotate in the plate
plane, the superfluid becomes ``spinful''.
By applying a stronger field, we can explore the spin polarized
A$_1$ phase with a substantial temperature window which
could be wider than the A phase window at $H=0$ 
under $H\sim10$ T\cite{sagan}.

In conclusion, we have proposed an elegant method to create 
the Majorana zero energy state by controlling the $l$-vector and $d$-vector
of the $^3$He-A phase via the boundary condition and external field.
The critical rotation velocity is accurately and realistically estimated
as a function of the system radius $R$ based on the established
GL functional form. The spectral properties of the stable
singular vortex are examined by solving the BdG equation for a 3D
Fermi surface, explicitly demonstrating that there exists a 
dispersionless Majorana state localized in a vortex core.
The utility of this 3D Majorana state for other fields, such as 
quantum computation, etc., is open for future study.

The authors thank O. Ishikawa and T. Ohmi for providing expertise on superfluid $^3$He.


\begin{thebibliography}{99}
\bibitem{GLparameter} The following GL parameters are 
used: $\alpha_0=3.81\times10^{50}[J^{-1}m^{-3}]$, $\beta_1=-3.75\times10^{99}[J^{-3}m^{-3}]$, $\beta_2=6.65\times10^{99}[J^{-3}m^{-3}]$, $\beta_3=6.56\times10^{99}[J^{-3}m^{-3}]$, $\beta_4=5.99\times10^{99}[J^{-3}m^{-3}]$, $\beta_5=-8.53\times10^{99}[J^{-3}m^{-3}]$, $2m_3/\hbar=9.51\times10^{7}[m^{-2}s].$
\end{thebibliography}
\end{document}